\documentclass{elsarticle}
\usepackage{ifpdf}
\usepackage{graphicx,amssymb,lineno}
\ifpdf
\usepackage[
  breaklinks=true,
  colorlinks=true,
  linkcolor=blue,anchorcolor=blue,
  citecolor=blue,filecolor=blue,
  menucolor=blue,pagecolor=blue,
  urlcolor=blue]{hyperref}
\fi

\makeatletter
\def\elsartstyle{
    \def\normalsize{\@setfontsize\normalsize\@xiipt{14.5}}
    \def\small{\@setfontsize\small\@xipt{13.6}}
    \let\footnotesize=\small
    \def\large{\@setfontsize\large\@xivpt{18}}
    \def\Large{\@setfontsize\Large\@xviipt{22}}
    \skip\@mpfootins = 18\p@ \@plus 2\p@
    \normalsize
} \@ifundefined{square}{}{} \makeatother

\begin{document}
\begin{frontmatter}
\title{Entangled coherent states under dissipation}
\author{F. Lastra,$^{3}$ G. Romero,$^{4}$ C. E. L\'{o}pez,$^{1,2}$ N. Zagury,$^{5}$
and J. C. Retamal$^{1,2}$}
\address{$^{1}$Departamento de
F\'{\i}sica, Universidad de Santiago de Chile, USACH, Casilla 307,
Santiago 2, Chile}
\address{$^{2}$Center for the Development of Nanoscience and
Nanotechnology, 9170124, Estaci\'on Central, Santiago, Chile}
\address{$^{3}$Facultad de F\'{\i}sica,
Pontificia Universidad Cat\'{o}lica de Chile, Casilla 306,
Santiago 22, Chile}
\address{$^{4}$Departamento de Qu\'{\i}mica F\'{\i}sica, Universidad del Pa\'{\i}s Vasco-Euskal Herriko Unibertsitatea, Apdo. 644, 48080 Bilbao, Spain}
\address{$^{5}$Instituto de F\'{\i}sica, Universidade Federal do Rio de
Janeiro, Caixa Postal 68528, Rio de Janeiro, RJ 21941-972, Brazil}
\begin{abstract}
We study the evolution of entangled coherent states of the two
quantized electromagnetic fields under dissipation. Characteristic
time scales for the decay of the negativity are found in the case
of  large values of the phase space distance among the states of
each mode. We also study how the entanglement emerges among the
reservoirs.
\end{abstract}
\begin{keyword}
Entanglement, Coherent States, Decoherence
 \PACS 03.67.Mn,03.65.Ud,03.65.Yz
\end{keyword}
\end{frontmatter}

\section{Introduction}

Entanglement of two or more systems is an important concept
playing a central role in quantum information science and quantum
computation~\cite{nielsen,ekert96}. It is an essential resource
for teleportation~\cite{bennett93}, quantum key
distribution~\cite{ekert91,gisin02} and controlled quantum
logic~\cite {knill01}, among others. The study of entangled states
of elementary quantum systems has become a fruitful field of
research
\cite{nielsen,ekert96,bennett93,ekert91,gisin02,knill01,tittel01,leibfried03,vandersypen04,raimond01,hanson06}.
In particular, it has been paid a great attention on studying the
behavior of quantum entanglement as a function of time when the
system is affected by the environment
\cite{DGCZ,prauzner,horodecki2001,Diosi,Dodd,eberly00,eberly1,Marcelo06,resto,almeida07,laurat07}.

The connection between decoherence and disentanglement has
received considerable attention in the last years. Whereas
decoherence of systems in contact with Markovian reservoirs decays
asymptotically, entanglement may disappear suddenly. This has been
recognized both in continuous variable
systems~\cite{DGCZ,prauzner,Zaba09} and  in finite dimensional
systems~\cite{horodecki2001,Diosi,Dodd,eberly00,eberly1,Marcelo06}.
In particular, entanglement among coherent states is a subject
that has  received attention by a number of authors
Ref.~\cite{sanders92,filip01,vanEnk01,jeong01,wang02,vanEnk03,vanEnk05}.

The purpose of this paper is to study entanglement properties of
superpositions of two-mode coherent states, in contact with a
dissipative reservoir paying particular attention to the general
features  of the dynamics of entanglement. When the phase
separation of the coherent states of each mode is large, we show,
in a large variety of numerical examples, that the entanglement
decays rapidly with a well defined characteristic time scale,
which is much smaller than the time, $ \tau$,  when the
entanglement completely disappears. This is in contrast to the
case of two two-level  systems, where the value of $\tau$ does have a
physical relevance. Analytical results are also found for
certain classes of states showing the same dynamical behavior. By
modeling the reservoirs through a set of linear oscillators we
also discuss how entanglement emerges among the reservoirs as the
system looses it.

This work is organized as follows: In Sec. II we present the
principal features of the entanglement dynamics under dissipation
when considering as the initial state a general superposition of
two coherent states for each mode. In Sec. III we study particular
situations where we may easily describe analytically the time
evolution of the negativity and the conditions for complete
disentanglement. In Sec. IV we consider explicitly the reservoir
degrees of freedom, in order to describe how the entanglement
emerges into the two reservoirs. Finally, in Sec. V we present our
concluding remarks.

\section{Disentanglement dynamics}
Consider the dynamics of two dissipative quantum modes, each
affected by its own environment. Such a situation can be
conveniently described  at zero temperature by the master equation
\begin{equation}  \label{master}
\hat{\dot\rho} = \frac{\Gamma_a}{2}\left ( 2 \hat a \hat\rho \hat
a^{\dagger}- \hat a^{\dagger} \hat a \hat \rho-\hat \rho \hat
a^{\dagger} \hat a \right )+ \frac{\Gamma_b}{2}\left ( 2 \hat b
\hat \rho \hat b^{\dagger}- \hat b^{\dagger} \hat b \hat \rho-\hat
\rho \hat b^{\dagger} \hat b \right ),
\end{equation}
where $\hat a,\hat a^{\dagger }$ and $\hat b,\hat b^{\dagger }$
are the annihilation and creation operators of the two modes.

Consider the class of initially entangled pure states of the form
\begin{equation}
\left\vert \Psi \right\rangle =\sum_{i,k=1,2}c_{i}^{k}\left\vert
\alpha _{i}\right\rangle \left\vert \beta _{k}\right\rangle ,
\label{initial2}
\end{equation}%
where $\left\vert \alpha _{_{1}}\right\rangle $ and $\left\vert
\alpha _{_{2}}\right\rangle $ ($\left\vert \beta
_{_{1}}\right\rangle $ and $\left\vert \beta _{_{2}}\right\rangle
$ ) are coherent states associated with mode $a$ (mode $b$) and
$c_{i}^{k}$ ($i,k=1,2$), are complex constants. An important
feature of the solution of Eq.~(\ref{master}) for $\hat{\rho}(t)$
is that, for each $t$ it can be written \emph{only} in terms of
the operators $|\alpha _{i}(t),\beta _{m}(t)\rangle $ $\langle
\alpha _{j}(t),\beta _{n}(t)|.$ In fact, it can be shown that this
solution is given by
\begin{equation}
\hat{\rho}(t)=\sum_{i,j,m,n=1,2}C_{i,j}^{m,n}(t)|\alpha
_{i}(t)\beta _{m}(t)\rangle \langle \alpha _{j}(t)\beta _{n}(t)|,
\label{rhot}
\end{equation}
where $\alpha _{i}(t)=\alpha _{i}e^{-\Gamma _{a}t/2},\,\beta
_{i}(t)=\beta _{i}e^{-\Gamma _{b}t/2},\ i=1,2,$ and

\begin{equation}
\begin{array}{l}
C_{i,j}^{m,n}(t)=  c_{i}^{m}(c_{j}^{n})^{\ast }\exp
{\{-(1-e^{-\Gamma _{a}t})(|\alpha _{i}-\alpha
_{j}|^{2}}+2i\mathrm{{Im}(\alpha _{i}\alpha _{j}^{\ast
}))/2\}\times } \\
\exp {\{-(1-e^{-\Gamma _{b}t})(|\beta _{m}-\beta
_{n}|^{2}}+2i\mathrm{{Im}({\beta _{m}\beta }_{n}^{\ast }))/2\}}.
\end{array}
\label{general}
\end{equation}
The state above could be conveniently written in an orthogonal
basis for each time $t.$ For example, we could use the basis

\begin{eqnarray}
|\xi_{\pm}(t)\rangle &=& N_{\xi_\pm}\bigg(|\xi _{1}(t)\rangle \pm
\frac{\langle \xi _{2}(t)|\xi _{1}(t)\rangle}{|\langle \xi
_{2}(t)|\xi _{1}(t)\rangle|}|\xi _{2}(t)\rangle \bigg),
\label{basis2q}
\end{eqnarray}
with $N_{\xi \pm}=1/\sqrt{2(1 \pm |\langle \xi _{2}(t)|\xi
_{1}(t)\rangle |)}$ and $\xi= \alpha, \beta$.

Although our system is a continuous one, it evolves in such way
that {\it at any time t} it can be described in the time dependent
subspace spanned by the vectors $\left|
\alpha_{\pm}(t)\right\rangle \otimes \left|
\beta_{\pm}(t)\right\rangle$. In this basis, we are allowed to
treat the whole system as an effective two-qubit
system~\cite{qubits} so that we can study entanglement properties
through the concurrence~\cite{wootters98}
\begin{equation}\label{concurrence}
C(t)={\rm{max}}\{0, (\lambda_1-\lambda_2-\lambda_3-\lambda_4)\},
\end{equation}
where $\lambda_i(t), i=1,2,3,4$ are the square root of
eigenvalues, in decreasing order, of the matrix
$M(t)=\rho(t)\sigma_{1y}\otimes\sigma_{2y}\rho(t)^*\sigma_{1y}\otimes\sigma_{2y},$
with all matrices been written in the  time dependent
4-dimensional basis [Eq.~(\ref{basis2q})].

The coherence of a superposition of a pair of coherent states
decays for short times with a time scale depending on the distance
between the superposed states and almost disappears much before
the characteristic time for energy dissipation. The problem we
want to address here is how entanglement is lost for a pair of
entangled coherent states and how it depends on the initial
conditions. For any state of the form of Eq.~(\ref{rhot}) there is
no simple analytical expression to describe how the entanglement
evolves. In this section we take a numerical approach to get some
insight in understanding the evolution of entanglement as a
function of distances between the superposed states.

For short times compared with $1/\Gamma _{a}(1/\Gamma _{b})$ the
matrix elements $C_{i,j}^{n,n}, i\ne j,$  ( $C_{i,i}^{m,n}, m\ne
n,$ )  decay exponentially with a time constant $\tau_a=\left[(
\Gamma_a|\alpha_1-\alpha_2|^2)/2 \right]^{-1}$ ( $\tau_b=\left[(
\Gamma_b|\beta_1 -\beta_2|^2)/2 \right]^{-1}$ ). $\tau_a $
($\tau_b$) are time scales associated to the loss of coherence
among the states  $|\alpha_1\rangle $ ($|\beta_1\rangle $) and
$|\alpha_2\rangle $($|\beta_2\rangle $  of mode $a$ $(b). $
Entanglement, as measured by the concurrence,
Eq.\ref{concurrence}, is a sum of non linear functions of the
matrix elements $C_{i,j}^{m,n}(t),$ and, depending on the initial
conditions, may even vanishes at finite times although  the matrix
elements $C_{i,j}^{m,n}(t),$ only vanish at infinity.   On
physical grounds, we expect that  entanglement evolution will
depend mostly on the matrix elements $C_{i,j}^{m,n}(t)$ with $
i\ne j, m\ne n.$ Therefore for values of $D^2=(|\alpha_1
-\alpha_2|^2+|\beta_1 -\beta_2|^2)/2\gg 1,$ and for times shorter
than $1/\Gamma_a$ and $1/\Gamma_b$  the concurrence should
initially drop exponentially in a  characteristic time $\tau =
1/[(\Gamma_a |\alpha_1 -\alpha_2|^2+\Gamma_b |\beta_1
-\beta_2|^2)/2].$ We show examples of this behavior by plotting, in
Fig.~\ref{lnCfixD}, $\ln C(t)/C(0)$ versus $t/\tau$ for
$\Gamma_a=\Gamma_b$,  for fixed values of $D^2=(|\alpha_1
-\alpha_2|^2+|\beta_1 -\beta_2|^2)/2$ and for values of the
constants $c_{i}^k$ chosen at random. The curves show evidence
that the entanglement initially does follow a single exponential
behavior. After a while some of the curves drop more rapidly
showing evidence of a non asymptotic decay of the concurrence,
suggesting that finite time complete disentanglement is present
for a variety of states. In the next section we study special
cases for which one may obtain simple analytical results for the
concurrence.

\section{Special cases}\label{secIII}

In this section we consider certain classes of states where we are
able to find simple analytical results for the concurrence,
showing explicitly the behavior of disentanglement suggested in
the last section.  In all cases we take, for simplicity,
$\Gamma_a=\Gamma_b=\Gamma $  and the amplitudes
$\alpha_1(\beta_1)$ and $\alpha_2(\beta_2)$ of the states in
Eq.~(\ref{initial2}) having the same phase. The density matrix in
Eq.~(\ref{rhot}) can then be written at each time $t$ in the
orthogonal basis defined by
\begin{eqnarray}
\mid \pm ,t\rangle _{a}&=& N_{a\pm }(t)(\mid \alpha _1({t})\rangle
\pm \mid
\alpha_2({t})\rangle ) \nonumber  \\
\mid \pm ,t\rangle _{b}&=& N_{b\pm }(t)(\mid \beta _1({t})\rangle
\pm \mid \beta _2({t})\rangle ),
\end{eqnarray}
where
\begin{eqnarray}
N_{a\pm }(t)&=&1/\sqrt{2(1 \pm e^{-\frac{1}{2}(|\alpha_1-\alpha_2|^2)e^{-\Gamma t}})}\nonumber\\
N_{b\pm }(t)&=&1/\sqrt{2(1 \pm e^{-\frac{1}{2}(|\beta_1-\beta_2|^2)e^{-\Gamma t}})}\nonumber\\
\end{eqnarray}
We also use, from now on the negativity, defined as  smallest
eigenvalue  of the partial transposition matrix
$\rho^{T_A}$~\cite{Peres96,Horodecki96}, as a measure of
entanglement.

As a first example we consider the state
\begin{equation}\label{equalweight}
|\Psi\rangle=\frac{1}{2}(|\alpha_{1}\beta_{1}\rangle+|\alpha_{1}\beta_{2}\rangle+|\alpha_{2}\beta_{1}\rangle-|\alpha_{2}\beta_{2}\rangle),
\end{equation}
which is a very high entangled state whenever
$d_{\alpha}^2=|\alpha_{1}-\alpha_{2}|^2/2 \gg 1$ and
$d_{\beta}^2=|\beta_1 -\beta_2|^2/2 \gg 1$. Within these
conditions  we obtain a  simple  form for the negativity:
\begin{equation}
\lambda_{-}=(1-e^{(e^{-\Gamma_a
t}-1)d_{\alpha}^2}-e^{(e^{-\Gamma_b
t}-1)d_{\beta}^2}-e^{(e^{-\Gamma_a t}-1)d_{\alpha}^2}
e^{(e^{-\Gamma_b t}-1)d_{\beta}^2}     )/4,
\end{equation}
so that entanglement decays with a superposition of exponentials
within a characteristic time $ 1/\left[
\Gamma(d_{\alpha}^2+d_{\beta}^2) \right]$ as suggested in section
II. Numerical results for the negativity are shown in
Fig.~\ref{figure4}, for $\Gamma_a=\Gamma_b=\Gamma$ and
$D^2=(d_{\alpha}^2+d_{\beta}^2 )/2$ equals to $0.5, 0.75, 1.5, 2$
and $5$.  As we observe in the figure, $\ln
\lambda_{-}(t)/\lambda_{-}(0)$ as a function of $\Gamma D^2 t$ has
a slope close to $-1,$  even for $D^2$  as small as 2.

Consider now that the system is initially prepared in  entangled
states of the form
\begin{equation}\label{state1}
\left| \Psi \right\rangle =c_{+,+}\left| +,0\right\rangle
_{a}\left| +,0\right\rangle _{b}+c_{-,-}\left| -,0\right\rangle
_{a}\left| -,0\right\rangle _{b},
\end{equation}
where, without loss of generality, $c_{+,+}$ and $c_{-,-}$ are
taken as real and positive constants. The density matrix written
in the basis $\{\mid \pm \pm ,t\rangle:=\mid \pm  ,t\rangle_a \otimes \mid \pm
,t\rangle_b \} $ has the form of an $X-{\rm matrix}$ as previously
considered for studying finite-time
disentanglement~\cite{eberly00,eberly1,Marcelo06}. In such case it
can be shown that the concurrence is related to the negativity,
$\lambda_{-},$ by the expression $C(t)=\textrm{max}\{0,-2\lambda
_{-}\}.$ For simplicity, let us assume equal distances for each
mode, $\left| \alpha_1 -\alpha_2 \right| =|\beta_1 -\beta_2 |$. The negativity can then be easily calculated and it is given by
\begin{equation}
\lambda _{-}=\left[ 1-e^{-D^2e^{-\Gamma t}}\right]\left[
A_2^2-A_1^2e^{-D^2(1-e^{-\Gamma t})}   \right], \label{lmenos}
\end{equation}
with $A_{1} =c_{+,+}N_{a+}(0)N_{b+}(0)+c_{-,-}N_{a-}(0)N_{b-}(0)$
and $A_{2} =c_{+,+}N_{a+}(0)N_{b+}(0)-c_{--}N_{a-}(0)N_{b-}(0)$.

Complete disentanglement occurs when the eigenvalue $\lambda _{-}$
becomes positive. From the expression above it is easy to see that
this happens in a time $t_{\rm{d}}^{(1)}$ given by:
\begin{equation}
t_{\rm{d}}^{(1)}=-\frac{1}{\Gamma }\ln \left( 1+\frac{2}{D
^{2}}\ln \left( \frac{\left| A_{2}\right| }{\left| A_{1}\right|
}\right) \right) .  \label{td}
\end{equation}
The condition for having finite disentanglement time is that $\exp
(-D ^{2}/2)<\left| A_{2}\right|/\left| A_{1}\right|,$ which can be
satisfied in two different cases. Firstly for $c_{-,-}/f>c_{+,+}$,
with $f=(1-e^{-D^2/2})/(1+e^{-D^2/2})$ for which there is always a
finite disentanglement time $t_{\rm{d}}^{(1)}$ if
$c_{-,-}>c_{+,+}$ . Secondly when $c_{+,+}>c_{-,-}/f$, where there
is $t_{\rm{d}}^{(1)}$ finite only if $c_{-,-}<c_{+,+}f^2$.

For large distances, $D \gg1$, we see from Eq.~(\ref{lmenos}) that
the concurrence drops exponentially with a characteristic time
$(D^2\Gamma)^{-1}, $ which is much shorter than the dissipation
time $\Gamma^{-1} , $ in agreement with the results suggested in
section II.

Although the loss of entanglement may happen at a finite time
$t_{\rm{d}}^{(1)}$, the concurrence, for $D > 1,$ is already small
at a much earlier time. A similar result was found in
Ref.~\cite{aolita08} when studying GHZ $N$ qubits states under
different independent reservoirs. Consider, for example, the
expression~(\ref{lmenos}) for the case of the initial state given by Eq.~(\ref{state1}). For
$D^2\gg 1,$ $\Gamma t_{\rm{d}}^{(1)}\ll 1$ we can easily show that
the negativity drops significantly to low values much before the
time $t_d$ when it completely disappears, independently of the
values of $c_{++},c_{--}$ and $D.$ In Fig.~\ref{lnvsR} we show the
behavior of the negativity for a fixed value of the ratio
$c_{++}/c_{--}$ and for several values of $D^2, $ as a function of
the normalized time $r=t/t_{\rm{d}}^{(1)}$. From that figure we
see that the negativity drops  to  very low values in a time less
than half the complete disentanglement time $t_d.$

For small distances, $D\ll1,$ Eq.~(\ref{lmenos}) can be
approximated to
\begin{equation}
\lambda _{-}\simeq c_{--}^2 e^{-\Gamma t}\left[
1-\frac{c_{++}}{c_{--}}-e^{-\Gamma t}\right]. \label{lmenos2}
\end{equation}
The interesting about this regime is that clearly shows two
different scales of disentanglement depending on the values of
$c_{++}$ and $c_{--}$. In particular, for an initial maximally
entangled state, that is, $c_{++}=c_{--}=1/\sqrt 2$,
$\lambda_{-}\simeq -(1/2)\exp{(-2\Gamma t})$ the entanglement
decays asymptotically with a rate $2\Gamma$. For $c_{-,-}>c_{+,+}$
entanglement vanish for time $t_d=-\Gamma ^{-1}\ln
(1-c_{+,+}/c_{-,-})$. For $c_{-,-}<c_{+,+}$ entanglement decays
always asymptotically. So that this case behaves in a similar way
to the two-qubit case in \cite{Marcelo06}.

Other situation of interest is the class of initial states having
the form:
\begin{equation}
\left| \Phi \right\rangle =c_{+,-}|+,0\rangle_a\left|
-,0\right\rangle _{b}+c_{-,+}\left| -,0\right\rangle _{a}\left|
+,0\right\rangle _{b}, \label{cmasmenos}
\end{equation}
where $c_{+,-}$, $c_{-,+}$ are taking as real constants.  The
dynamics of this state leads also to an X density matrix. In a
similar way as for the previous case, we found that the time for
disentanglement is given by:
\begin{equation}
t_{\rm{d}}^{(2)}=-\frac{1}{\Gamma }\ln \left( 1+\frac{2}{D^{2}}\ln
\left|\frac{(c_{+,-}-c_{-,+})}{(c_{+,-}+c_{-,+})}\right | \right).
\end{equation}
The only condition required for having $t_{\rm{d}}^{(2)}$ finite
is that: 
\begin{equation}
\exp (-\frac{1}{2}D^{2})<\left| (c_{+,-}-c_{-,+}) /
(c_{+,-}+c_{-,+} )\right|.
\end{equation}

\section{Birth of entanglement among the reservoirs}

In the above sections, we have studied disentanglement dynamics
for some special cases of entangled coherent states, paying
attention only to the system of the modes $a$ and $b$.
However, when including degrees of freedom of both reservoirs, a
deeper understanding of entanglement dynamics can be obtained.
This can be carried out by describing the system-reservoir
dynamics per each mode with a Hamiltonian of the form~\cite{Lopez08}
\begin{equation}
H=\hbar \omega a^\dagger a +\hbar \sum_{k=1}^{N}\omega_k
c^{\dagger}_kc_k+\hbar \sum_{k=1}^{N}g_k(
ac^{\dagger}_k+c_ka^{\dagger}).
\end{equation}
In such case, it is not difficult to show that when the mode $a$
is initially in a coherent state and the reservoir is in the
collective vacuum state
$|\bar{0}\rangle_{r}=\prod^{N}_{k=1}|0_k\rangle_{r}$, the mode $a$ and its reservoir evolve into a product of coherent states of the form
$\mid \alpha _t \rangle | {\bar{\alpha}_t}\rangle_{r}=\mid \alpha
F(t) \rangle \prod^{N}_{k=1} | f_k (t) \alpha\rangle_{r}.$ In  the 
Born-Markov approximation  $F(t)=e^{-\Gamma_a t/2}$ and $\sum_k
f_k (t)^2=1-e^{-\Gamma_a t}.$

To illustrate how entanglement emerges into the reservoirs, let us
consider the initial state of Eq.~(\ref{state1}), but now
including the states for both reservoirs
\begin{equation}
\left| \Psi \right\rangle =\big(c_{+,+}\left| +,0\right\rangle
_{a}\left| +,0\right\rangle _{b}+c_{-,-}\left|
-,0\right\rangle _{a}\left| -,0\right\rangle
_{b}\big)|\bar{0}\rangle_{r_a}|\bar{0}\rangle_{r_b},
\label{statecr}
\end{equation}
where $c_{+,+}$ and $c_{-,-}$ are real and positive constants.
Note that $a(b)$ and $r_{a(b)}$ stand for mode $a(b)$ and
reservoir associated to mode $a(b)$, respectively.
Since the evolution of each mode and its reservoir may be treated
independently, the complete density matrix at time $t$ can be
obtained through the evolution of the states
$|\pm,0\rangle_a|\bar{0}\rangle_{r_a}=N_{a\pm}(0)\big(|\alpha\rangle
\pm |-\alpha\rangle \big)|\bar{0}\rangle_{r_a}$ and
$|\pm,0\rangle_b|\bar{0}\rangle_{r_{b}}=N_{b\pm}(0)\big(|\beta\rangle
\pm |-\beta\rangle \big)|\bar{0}\rangle_{r_b}.$   

Let $c$ and $r_c$ stand for either mode $a$ or $b$ and for their own reservoirs. It is possible to
show that the initial states $|+,0\rangle_c|\bar{0}\rangle_{r_c}$ and $|-,0\rangle_c|\bar{0}\rangle_{r_c}$ evolve at the time $t$ to the states\begin{eqnarray}
|\phi_{+,t}\rangle_{rc}&=&\frac{N_{c+}(0)B_{c+}(t)}{2N_{c+}(t)}|+,t\rangle_c|E_{+,t}\rangle_{r_c}+\frac{N_{c+}(0)B_{c-}(t)}{2N_{c-}(t)}|-,t\rangle_c|E_{-,t}\rangle_{r_c}
\nonumber
\\
|\phi_{-,t}\rangle_{rc}&=&\frac{N_{c-}(0)B_{c-}(t)}{2N_{c+}(t)}|+,t\rangle_c|E_{-,t}\rangle_{r_c}+\frac{N_{c-}(0)B_{c+}(t)}{2N_{c-}(t)}|-,t\rangle_c|E_{+,t}\rangle_{r_c} \, ,
\nonumber \\
\end{eqnarray}
respectively. The states $|E_{\pm,t}\rangle_{r_c}$ in the equations above are defined as
$|E_{\pm,t}\rangle_{r_a}=\big(|\bar{{\bf \alpha}_t}\rangle_{r_a}
\pm|-\bar{{\bf \alpha}_t}\rangle_{r_a}\big)/B_{a\pm}(t)$ and
$|E_{\pm,t}\rangle_{r_b}=\big(|\bar{{\bf \beta}_t}\rangle_{r_a}
\pm|-\bar{{\bf \beta}_t}\rangle_{r_a}\big)/B_{b\pm}(t)$ and  form a two-dimensional orthogonal basis for each reservoir,  and normalization terms read
\begin{eqnarray}
B_{a\pm}(t)&=&\sqrt{2(1\pm e^{-{2}|\alpha|^2(1-e^{-\Gamma_a t})})}\, , \nonumber \\
B_{b\pm}(t)&=&\sqrt{2(1\pm e^{-{2}|\beta|^2(1-e^{-\Gamma_b t})})}.
\end{eqnarray}

Let us now study the joint evolution of the two modes coupled to
their reservoirs. In the above description, the initial
state~(\ref{statecr}) evolves to
\begin{equation}\label{purewhole}
|\Psi\rangle_t=c_{++}|\phi_{+,t}\rangle_{a r_{a}}|\phi_{+,t}\rangle_{b r_{b}}+c_{--}|\phi_{-,t}\rangle_{a r_{a}}|\phi_{-,t}\rangle_{b r_{b}}.
\end{equation}
This is a pure state representing the whole system. As $t$ increases, there is a transfer of the correlationsÊof the system, both classical and quantum, to the reservoirs. Notice thatÊ
there is no memory effects involved in this transfer. Information about the reservoir-reservoir dynamics are obtained by tracing out cavities degrees of freedom, as well as the cavity modes evolution are obtained by tracing out theÊstates of the reservoirs.

As before, let us consider equal distances for each mode,
$D^2=|\alpha_1-\alpha_2|^2=|\beta_1-\beta_2|^2$. The dynamics
given by Eq.~(\ref{purewhole}) leads us to an $X-{\rm matrix}$ for
the reservoir-reservoir subsystem,  which is complementary to the
density matrix of the subsystem of the two modes $a$ and $b$~\cite{Lopez08}. As a
consequence, we can study entanglement dynamics associated to
external degrees of freedom.

When studying entanglement through the partial transposition
matrix, we obtain a complementary result of~(\ref{lmenos}) giving
us information about the birth of entanglement in the
reservoir-reservoir subsystem. In this case, we obtain the
following negative eigenvalue
\begin{equation}
\lambda _{-}=\left[ 1-e^{-D^2(1-e^{-\Gamma t})}\right]\left[
A_2^2-A_1^2e^{-D^2e^{-\Gamma t}}   \right], \label{lmenos3}
\end{equation}
from which we can derive the time for birth of entanglement in
reservoirs, that is
\begin{equation}\label{tb}
t_{b}=-\frac{1}{\Gamma}\ln\bigg(\frac{2}{D^2}\ln\bigg(\frac{|A_1|}{|A_2|}\bigg)\bigg).
\end{equation}
Note that Eq.~(\ref{lmenos3}) can be obtained from
Eq.~(\ref{lmenos}) by exchanging the time-dependent functions,
$e^{-\Gamma t}$ and $1-e^{-\Gamma t}$. Also
\begin{equation}\label{tbd}
e^{-\Gamma t_d}+e^{-\Gamma t_b}=1
\end{equation}

This result immediately shows that if the entanglement among the
two modes persists all the time, the entanglement among the
reservoirs starts growing at $t=0. $ Eqs.~(\ref{td})
and~(\ref{tb}) shows that in this case the entanglement among the
reservoirs never dies and its negativity reaches at infinity the
initial value of the negativity among the two modes $a$ and $b.$
Eq.~(\ref{tbd}), together with Eqs.~(\ref{td}) and~(\ref{tb}) also
shows that the time when entanglement starts among the two
reservoirs can be either smaller, greater or equal to the time
when entanglement among the two modes completely
disappears. This result is analogous to the one obtained in Ref.~\cite{Lopez08}.

Figure~\ref{figure5} shows the entanglement evolution when $t_b >t_d$, considering two distances in phase space $D^2=0.4$ (Fig.~\ref{figure5}(a)) and $D^2=16$ (Fig.~\ref{figure5}(b)). Here there is a time window where no entanglement exist either among the two modes or among the two reservoirs. However, we observe that entanglement does exist in others bipartite partitions of the whole system, since each mode entangles with the reservoir of the other mode. When plotting the concurrences $\mathcal{C}_{ab}$, $\mathcal{C}_{r_ar_b}$, $\mathcal{C}_{ar_a}$ and $\mathcal{C}_{ar_b}$, we notice that in the region where there is no entanglement either among the modes $a$ and $b$ or among the two reservoirs, that is, $\mathcal{C}_{ab}=\mathcal{C}_{r_ar_b}=0$, entanglement between a mode and its corresponding reservoir reaches its maximum value. Although this behavior is presented in both cases, the correlations are sensitive to any change of parameter $D^2$, as shown in Fig.~\ref{figure5}(b) where the concurrence between mode $a$ and reservoir $r_b$ is very small. Moreover for small distances ($D^2=|2\alpha|^2$), which also correspond to a small average number of photons, we reproduce the entanglement dynamics associated to the two-qubit case studied in Ref.~\cite{Lopez08}. Note that since we are considering equal distances and decay rates, both mode-reservoir subsystems evolves identically implying that the concurrences $\mathcal{C}_{ar_a}=\mathcal{C}_{br_b}$ and $\mathcal{C}_{ar_b}=\mathcal{C}_{br_a}$. 

A particular result happens when $|A_1/A_2|=e^{D^2/4},$ that is
$c_{+,+}/c_{-,-}= (1+e^{-D^2/2})/(1-e^{-D^2/4})^2$. In this case,
the times when the modes completely disentangles and the time when
the reservoirs starts to be entangled are equal and occurs at
$t_b=t_d=\ln (2)/\Gamma,$ a result which in independent of the
distances in phase space for both modes.

\section{Conclusion}
In conclusion, we have studied the evolution, under dissipation, of the entanglement among two modes of the electromagnetic field for certain class of initially entangled coherent states. At zero temperature,
the density matrix can always be written, for each $t,$ in a
finite orthogonal basis, which allow us to describe the
entanglement evolution as that happening in a discrete system.
Asymptotic entanglement decay as well as finite time
disentanglement arises depending on the initial conditions for the
bipartite system and on the distances in phase space among the
components of each mode. Typically, the robustness of entanglement
decays in a time scale proportional to
$2/(\Gamma_a|\alpha_1-\alpha_2|^2+\Gamma_b|\beta_1-\beta_2|^2)$,
the inverse of a weighted average of the distances in phase space for
each mode, whenever these distances are much greater than one. Both numerical and analytical results are presented.

We have also presented an investigation concerning the entanglement
generation among the reservoirs. By modeling the reservoir and its
interaction with the system, it is possible to study death of
entanglement among both modes of the field and the birth of entanglement in
reservoir-reservoir subsystem, respectively. In addition to the
dependence on initial amplitudes, the characteristic times for death
and birth of entanglement depends on the distances among coherent
states. An interesting result shows that for specific conditions in
the initial amplitudes for a given distance among coherent states,
both birth and death of entanglement at finite times occur
simultaneously at a time which does not depend on the distances of
each mode.

\bigskip
\bigskip

\noindent{\textbf{Acknowledgements}}

\bigskip
\bigskip

F.L. acknowledges Fondecyt 3085030, G.R. acknowledges Juan de la Cierva Program, C.E.L. acknowledges Fondecyt
11070244 and PBCT-CONICYT PSD54, J.C.R. acknowledges Fondecyt
1070157. N.Z. acknowledges the support from CNPq and FAPERJ.
C.E.L. and J.C.R. acknowledge support from Financiamiento Basal
para Centros Cient\'icos y Tecnol\'ogicos de Excelencia.

\newpage

\begin{figure}[t]\label{lnCfixD}
\begin{center}
\includegraphics[width=60mm]{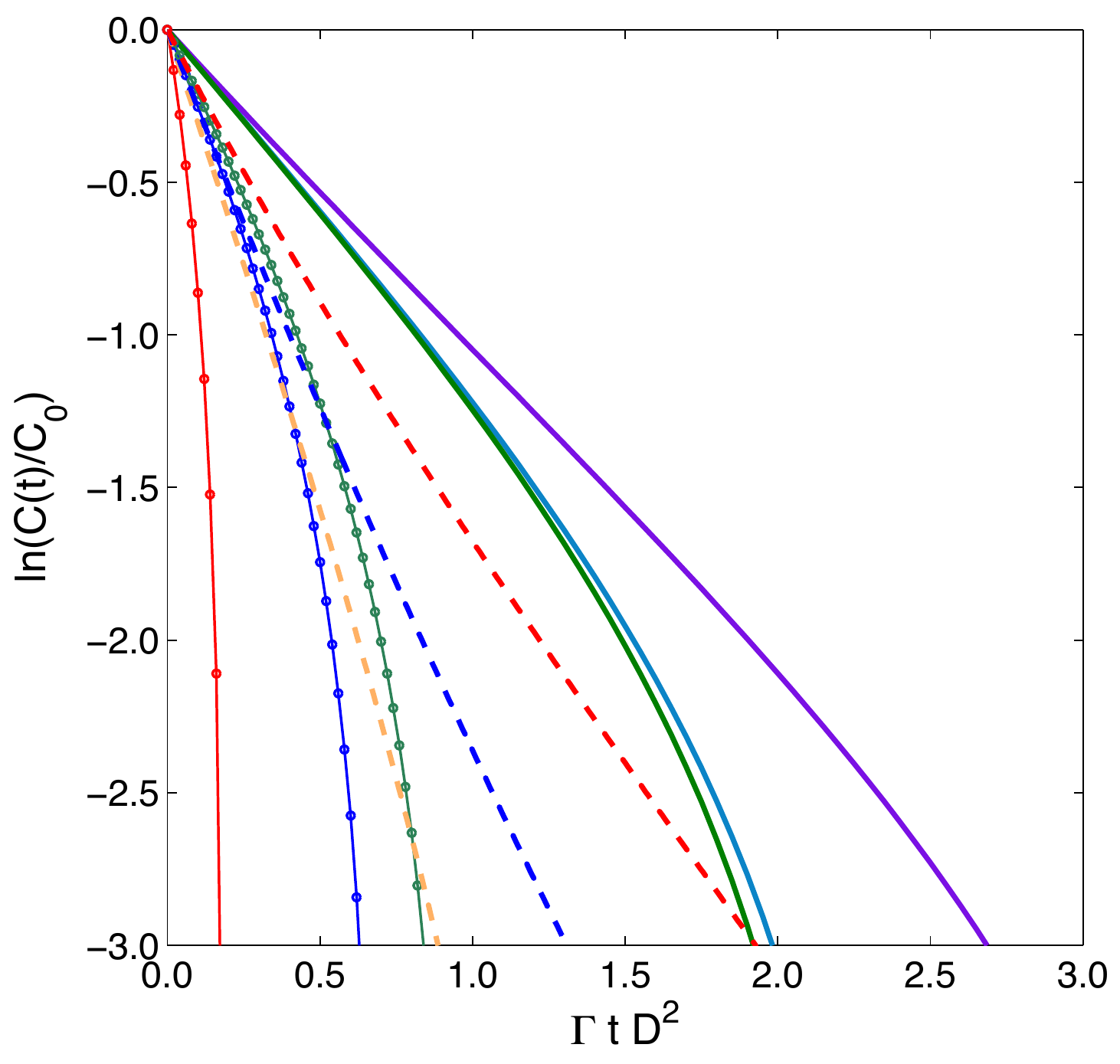}
\end{center}
\caption{$\ln C(t)/C(0)$ as a function of $t/\tau,$ for $D^2=0.75$
(line  and circles), $D^2=2$ ( dashed line) and $D^2=5$ (solid
line ). For each value of $D^2$, three sets of coefficients
$c_{i}^k$ that define the initial state~(\ref{initial2}) were
chosen at random. $\Gamma_{a}=\Gamma_{b}=\Gamma$}
 \label{lnCfixD}
\end{figure}

\begin{figure}[t]
\begin{center}
\includegraphics[width=60mm]{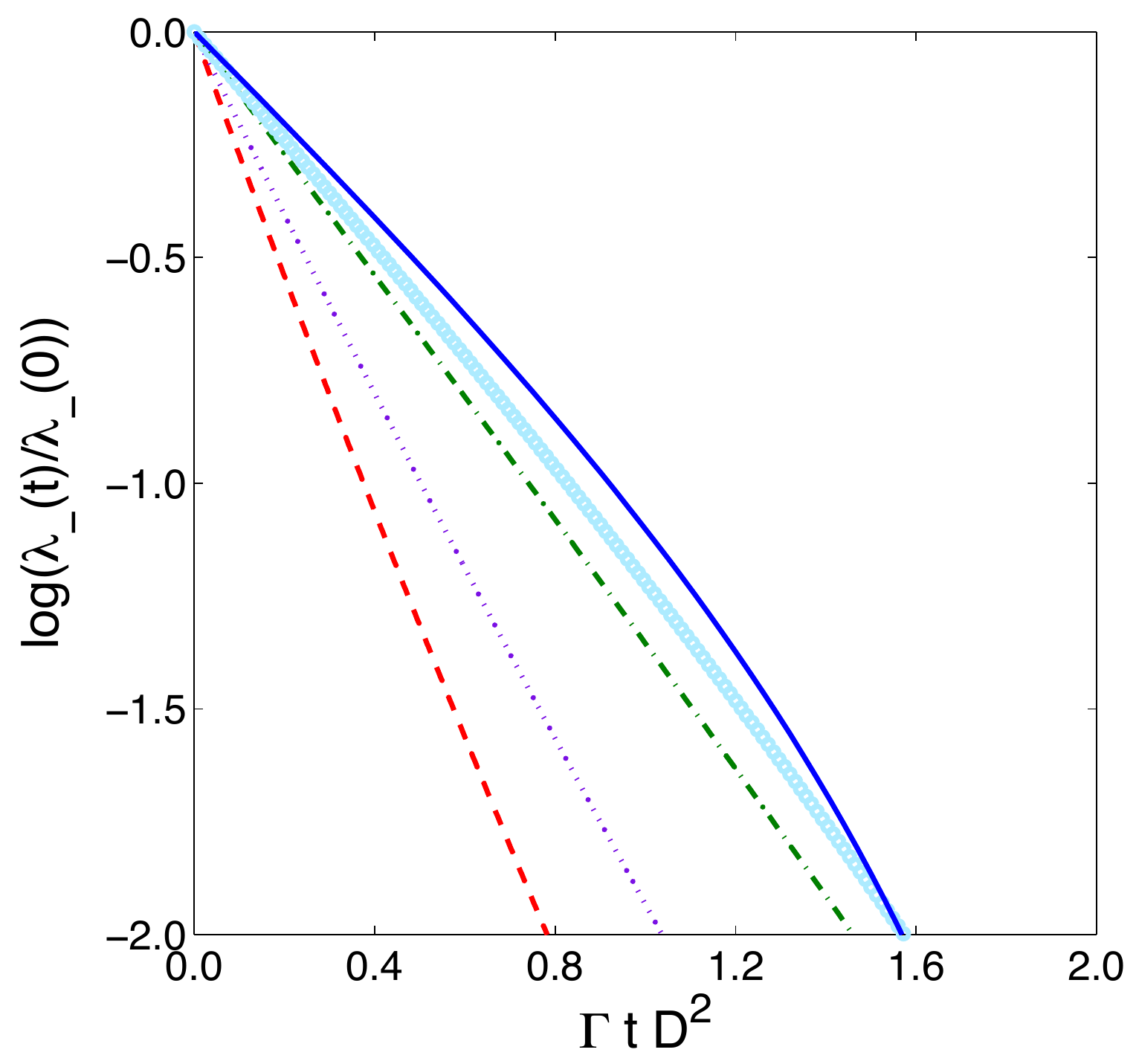}
\end{center}
\caption{$\ln \lambda(t)/\lambda(0)$ as a function of the
renormalized time $t \left[(
\Gamma_a|\alpha_1-\alpha_2|^2+\Gamma_b|\beta_1 -\beta_2|^2)/2
\right],$ for the state given by Eq.~\ref{equalweight}. $D^2=0.5$
(dashed line), $D^2=0.75$ (dotted line), $D^2=1.5$ (dot-dashed
line), $D^2=2$ (circles), $D^2=5$ (solid line).
$\Gamma_{a}=\Gamma_{b}=\Gamma$} \label{figure4}
\end{figure}

\begin{figure}[t]
\begin{center}
\includegraphics[width=60mm]{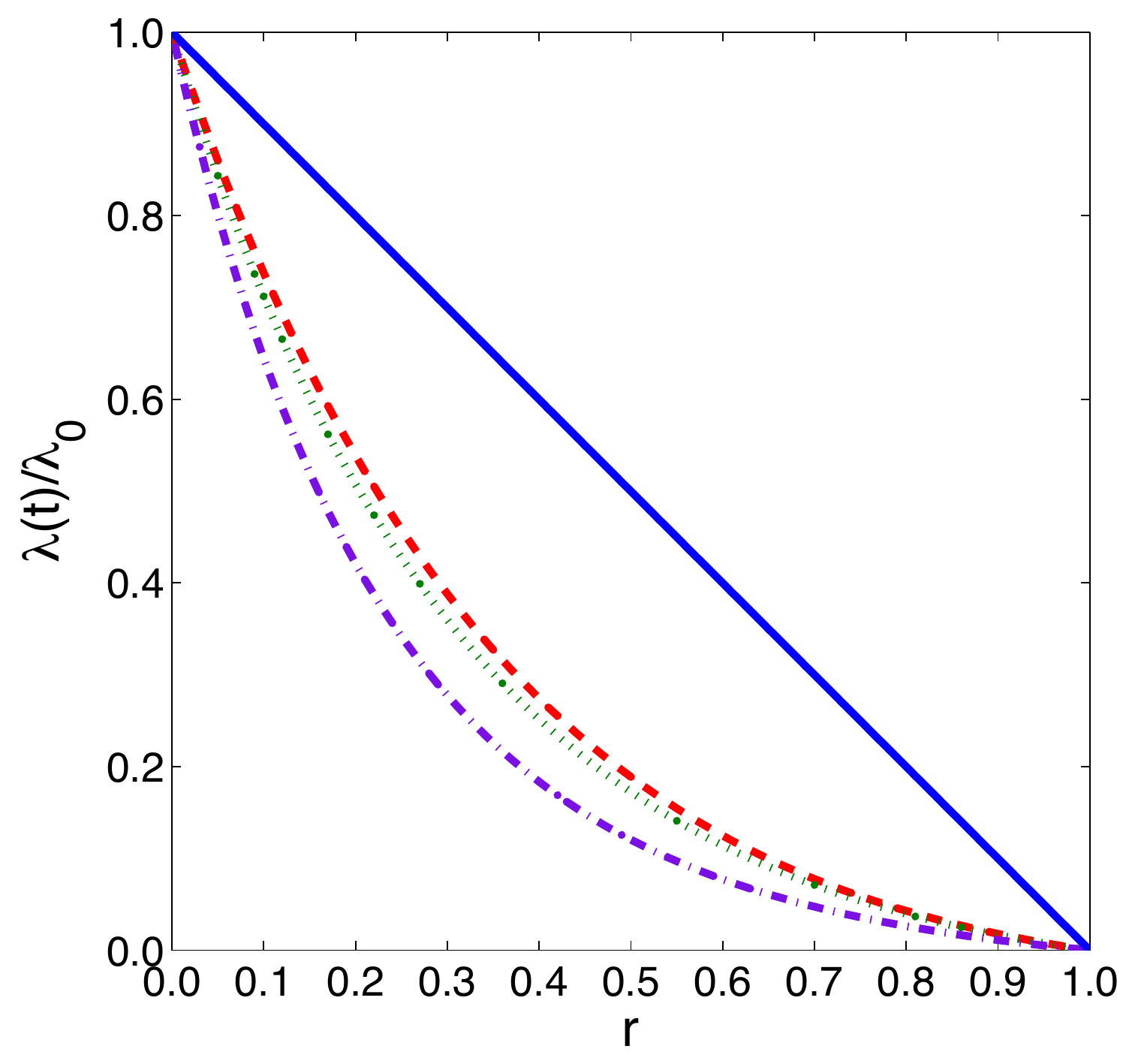}
\end{center}
\caption{Ratio of the negativity at time $t$ and at the initial
time as a function of the normalized time $r=t/t_{\rm{d}}^{(1)}$,
for fixed $|c_{--}/c_{++}|=\sqrt 2$ and distances $D^2=0.5$
(dashed line), $D^2=2$ (dotted line) and $D^2=7.5$ (dot-dashed
line). It is also shown, for comparison, the line
$1-r$.\label{lnvsR}}
\end{figure}

\begin{figure}[t]
\begin{center}
\includegraphics[width=12cm]{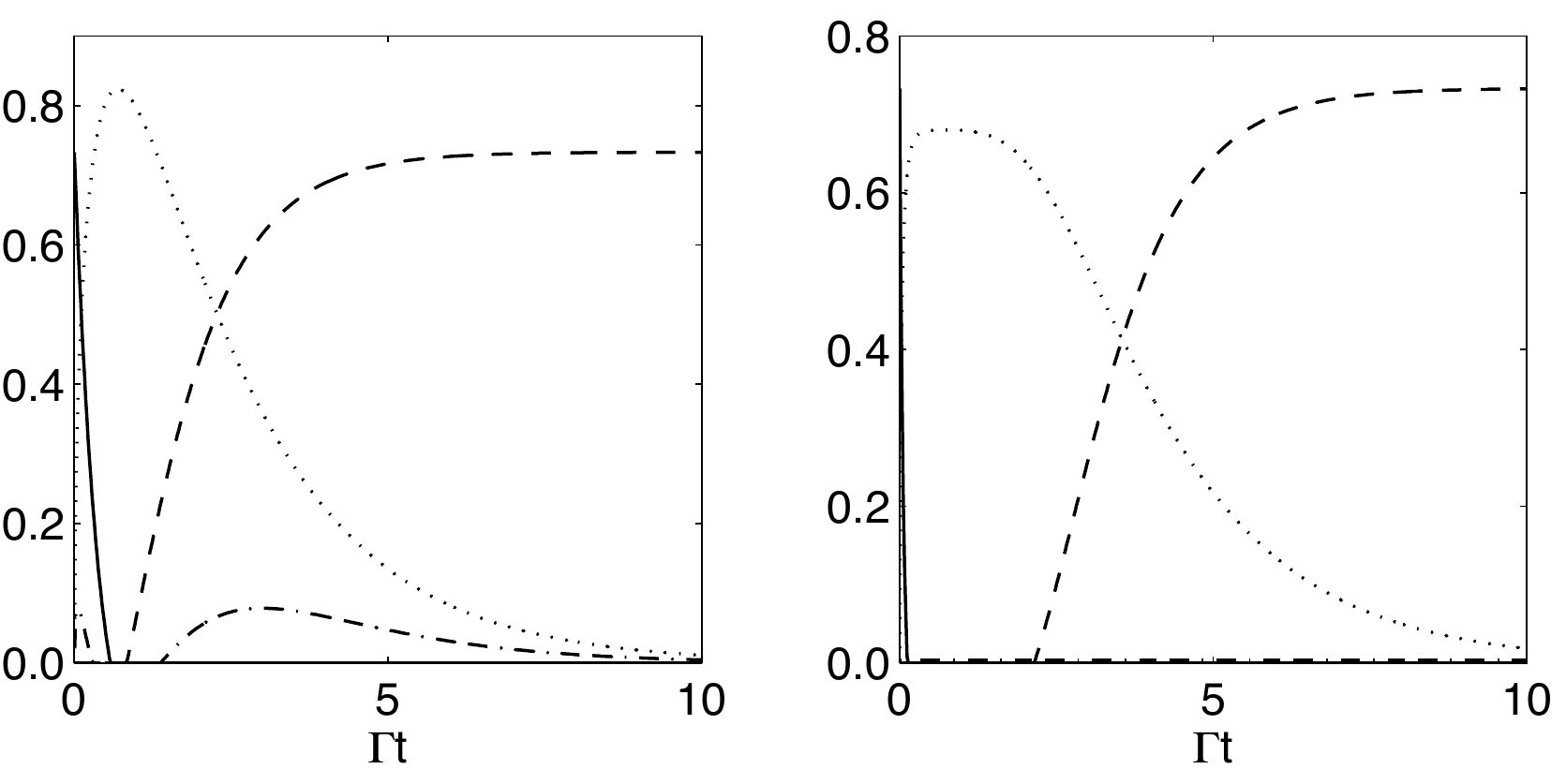}
\end{center}
\caption{Evolution of concurrence for different bipartite
partitions. $\mathcal{C}_{c_ac_b}$ (solid line),
$\mathcal{C}_{r_ar_b}$ (dashed line), $\mathcal{C}_{c_ar_a}$
(dotted line), and $\mathcal{C}_{c_ar_b}$ (dot-dashed line). (a) considering $D^2=0.4$ and (b) for $D^2=16$. In
both simulations we considered equal decay rates $\Gamma_{a}=\Gamma_{b}=\Gamma$, and 
amplitudes are $c_{+,+}=0.4$, $c_{-,-}=\sqrt{1-c_{+,+}^2}$.} \label{figure5}
\end{figure}


\begin{thebibliography}{9}
\bibitem{qubits} A two-qubit treatment of such a systems has also been considered in~\cite{jeong01}.
\end{thebibliography}
\end{document}